\documentclass[twocolumn,superscriptaddress,prl]{revtex4-1} 
\usepackage[dvipdfmx]{graphicx}
\usepackage{amsmath}
\usepackage{color}

\begin{document}

\title{Machine learning with controllable quantum dynamics \\
of a nuclear spin ensemble in a solid}

\author{Makoto Negoro}
\affiliation{Graduate School of Engineering Science, Osaka University, Toyonaka, Osaka 560-8531, Japan.}
\affiliation{JST, PRESTO, Kawaguchi, Saitama 332-0012, Japan}
\email{negoro@ee.es.osaka-u.ac.jp}
\author{Kosuke Mitarai}
\affiliation{Graduate School of Engineering Science, Osaka University, Toyonaka, Osaka 560-8531, Japan.}
\author{Keisuke Fujii}
\affiliation{Graduate School of Science, Kyoto University, Sakyo-ku, Kyoto 606-8302, Japan.}
\affiliation{JST, PRESTO, Kawaguchi, Saitama 332-0012, Japan}
\author{Kohei Nakajima}
\affiliation{Graduate School of Information Science and Technology, The University of Tokyo, Bunkyo-ku, Tokyo 113-8656, Japan.}
\affiliation{JST, PRESTO, Kawaguchi, Saitama 332-0012, Japan}
\author{Masahiro Kitagawa}
\affiliation{Graduate School of Engineering Science, Osaka University, Toyonaka, Osaka 560-8531, Japan.}

\date{\today}

\begin{abstract}
We experimentally demonstrate quantum machine learning using NMR based on a framework of quantum reservoir computing.
Reservoir computing is for exploiting natural nonlinear dynamics with large degrees of freedom, which is called a reservoir, for a machine learning purpose.
Here we propose a concrete physical implementation of a quantum reservoir using controllable dynamics of a nuclear spin ensemble in a molecular solid.
In this implementation, we demonstrate learning of nonlinear functions with binary or continuous variable inputs with low mean squared errors.
Our implementation and demonstration paves a road toward exploiting quantum computational supremacy in NMR ensemble systems for information processing with reachable technologies.
\end{abstract}

\maketitle

Since Shor has developed a quantum algorithm with exponential speedup, quantum computing experiments have been extensively carried out with various quantum systems such as trapped ions, photons, semiconductor quantum dots, superconducting circuits, and so on~\cite{NielsenChuang,Ladd10}.
Among them, ensemble NMR quantum computers have played a significant role as a testbed for quantum algorithms~\cite{Cory00,Jones00}.
In this approach, nuclear spins in a molecule in a bulk sample are used as qubits.
The sample has a macroscopic ensemble of identical copies of the qubit system.
We can control the qubit system with NMR and obtain an expectation value of an observable of the system directly from the NMR signal.
The NMR quantum computer has succeeded in the implementation of Shor's algorithm using 7 qubits~\cite{Vandersypen01}, quantum simulation using 7 qubits~\cite{Negrevergne05}, and quantum machine learning using 4 qubits~\cite{Biamonte17,Li15}.

However, the pseudo-initialization technique used in these implementations is not scalable and this fact cancels the exponential speedup of the quantum algorithms.
The state with non-initialized but partially-polarized spins in these implementations is inevitably separable~\cite{Braunstein98}.
In order to efficiently solve problems in BQP (bounded error quantum polynomial time) class like factoring, the scalable initialization of nuclear spin qubits is required.
A class of problems that can be solved scalably by quantum computers with partially-polarized states is called DQC1 (deterministic quantum computation with 1 qubit)~\cite{Knill98}.
Recently, it has been theoretically proven that DQC1 is unsimulatable with classical computers under stable complexity conjectures~\cite{Morimae14}, that is, {\it quantum computational supremacy}~\cite{Preskill12}.
The quantum simulation experiments~\cite{Alvalez15} have demonstrated that the controllable dynamics of partially-polarized nuclear spins in molecular solids can be really complex as more than 1000 spins are quantum mechanically correlated.

Here we experimentally demonstrate how to exploit such complex quantum dynamics of a nuclear spin ensemble for information processing, specifically for machine learning.
To this end, we employ quantum reservoir framework~\cite{Fujii17}.
Reservoir computing is a framework for exploiting a nonlinear dynamics with large degrees of freedom, called a {\it reservoir} here, for machine learning as it is~\cite{Maass02,Jaeger04,Versteraeten07}.
It has been first proposed as echo state network or liquid state machine~\cite{Maass02,Jaeger04}, where a conventional neural network is used as a reservoir, but its internal dynamics is randomly prefixed and only the linear readout weights are optimized to learn a nonlinear task.
This black box property allows us to use actual physical systems using photonics~\cite{Vandoorne14}, spintronics~\cite{Torrejon17}, or soft robotics~\cite{Nakajima15}, as well as qubits~\cite{Fujii17}.

Here we propose a concrete physical implementation of a {\it quantum reservoir} using {\it globally controlled} nuclear spins with a partially-polarized state in a molecular solid.
Some qubits are said to be globally controlled when they are manipulated simultaneously and equivalently.
It has been shown that even universal quantum circuits can be implemented with global control~\cite{Lloyd93,Benjamin03}.
In principle, classical computers cannot simulate the globally controlled circuits with partially-polarized states as well as initialized states.
In this work, we used isotopically-labelled {\it l}-alanine consisting of four $^1$H spins and one $^{13}$C spin as a reservoir, which is diluted into a single crystal of $^2$H$_7$-{\it l}-alanine.
As a proof-of-principle experimental demonstration, we perform non-temporal tasks using the natural quantum dynamics of the quantum reservoir, which we also call {\it quantum extreme learning machine}~\cite{Butcher13}.

The quantum circuit of the physical implementation is shown in Fig.~1A.
The experiment was carried out in a static magnetic field of 11.7~T at room temperature (See also Supplementary Information).
In order to implement the quantum circuit, we applied continuous RF irradiations resonating with $^1$H and $^{13}$C spins, as shown in Fig.~1B.
During the time evolution of the dipolar interactions with the reservoir spins, $U$, we applied so-called Hartmann-Hahn irradiation, where both of the Rabi frequencies were 83 kHz.
The purpose of the strong irradiation is to decouple the dipolar interactions between the reservoir spins and environmental $^2$H or $^{14}$N spins ($<$10 kHz).
Accordingly, the ensemble reservoir is well isolated.
The effective Hamiltonian in the interaction frame rotating around the static and RF fields is written as,
\begin{equation}
H = \sum_{i} d_{i,C} (I_X^i I_X^C +I_Y^i I_Y^C) + \sum_{i} d_{i,j} (2I_Z^i I_Z^j - I_Y^i I_Y^j - I_X^i I_X^j), \ \ \ 
\end{equation}
where $I_a^i$ and $I_a^C\ (a=X,$ $Y,$ and $Z)$ are the spin operators of the $i$-th $^1$H and the $^{13}$C.
$d_{i,C}$ and $d_{i,j}$ are the dipolar coupling constants between the $i$-th $^1$H spin and the $^{13}C$ spin and between the $i$-th and $j$-th $^1$H spins, respectively.
The time evolution $U=e^{-iHt}$ is complex ensemble dynamics where the $Z$ components are exchanged among all the spins in the reservoir.

\begin{figure}[tbp]
\center{\includegraphics[width=\linewidth]{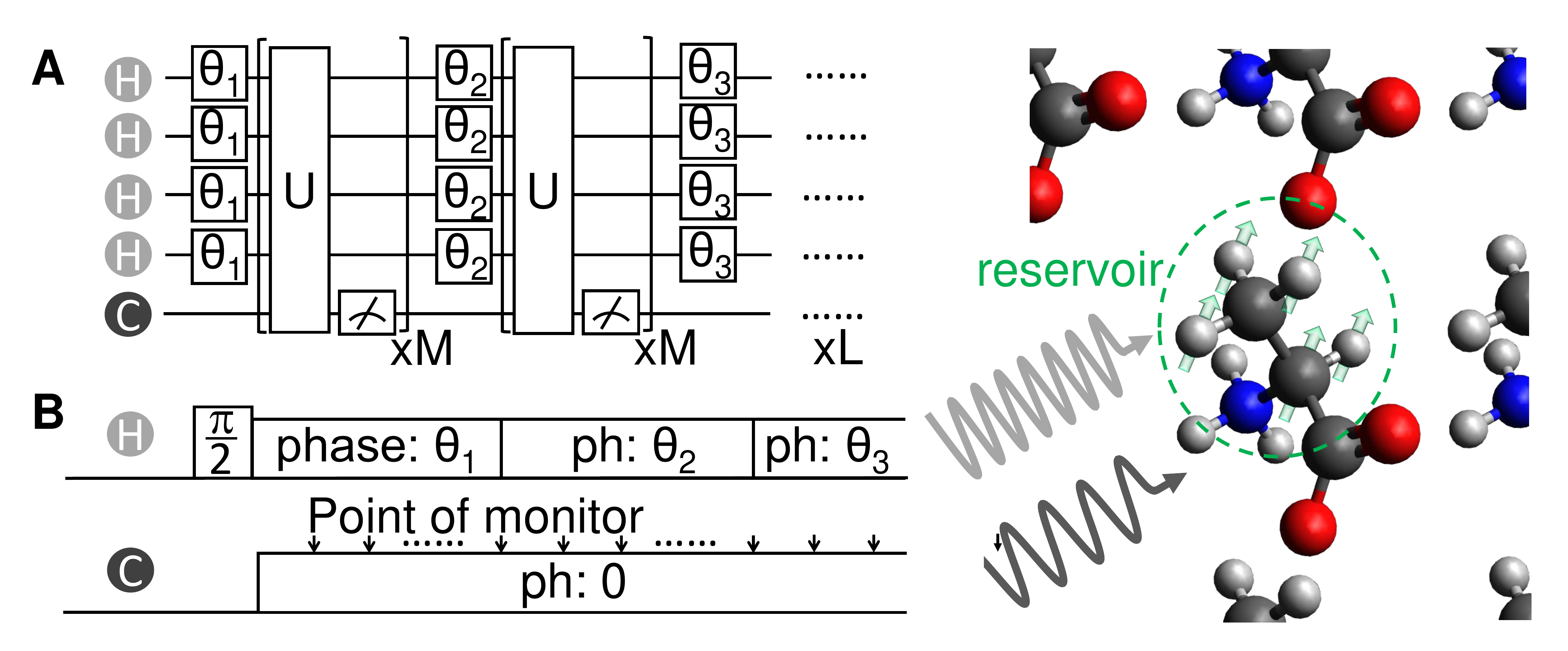}}
\caption{\label{fig1}
\small
Quantum circuit (A) and pulse sequence (B) of the proposed implementation for learning with a quantum reservoir.
}
\end{figure}

In order to control the reservoir dynamics, we employed the global rotation in the interaction frame by switching the phase of the RF field for the $^1$H spins.
This implementation allows us to utilize $^1$H spins in solids, which typically exist in large numbers in a molecule but have not been previously utilized as computational resources.
Typical dynamics of the reservoir are shown in Fig.~2.
The initial state before the Hartmann-Hahn irradiation is $\rho = \epsilon \sum_i {I_Z^i}+{\bf 1}/2^5$ in the interaction frame, where $\epsilon$ is the polarization of the $^1$H spins at thermal equilibrium.

\begin{figure}[tbp]
\center{\includegraphics[width=\linewidth]{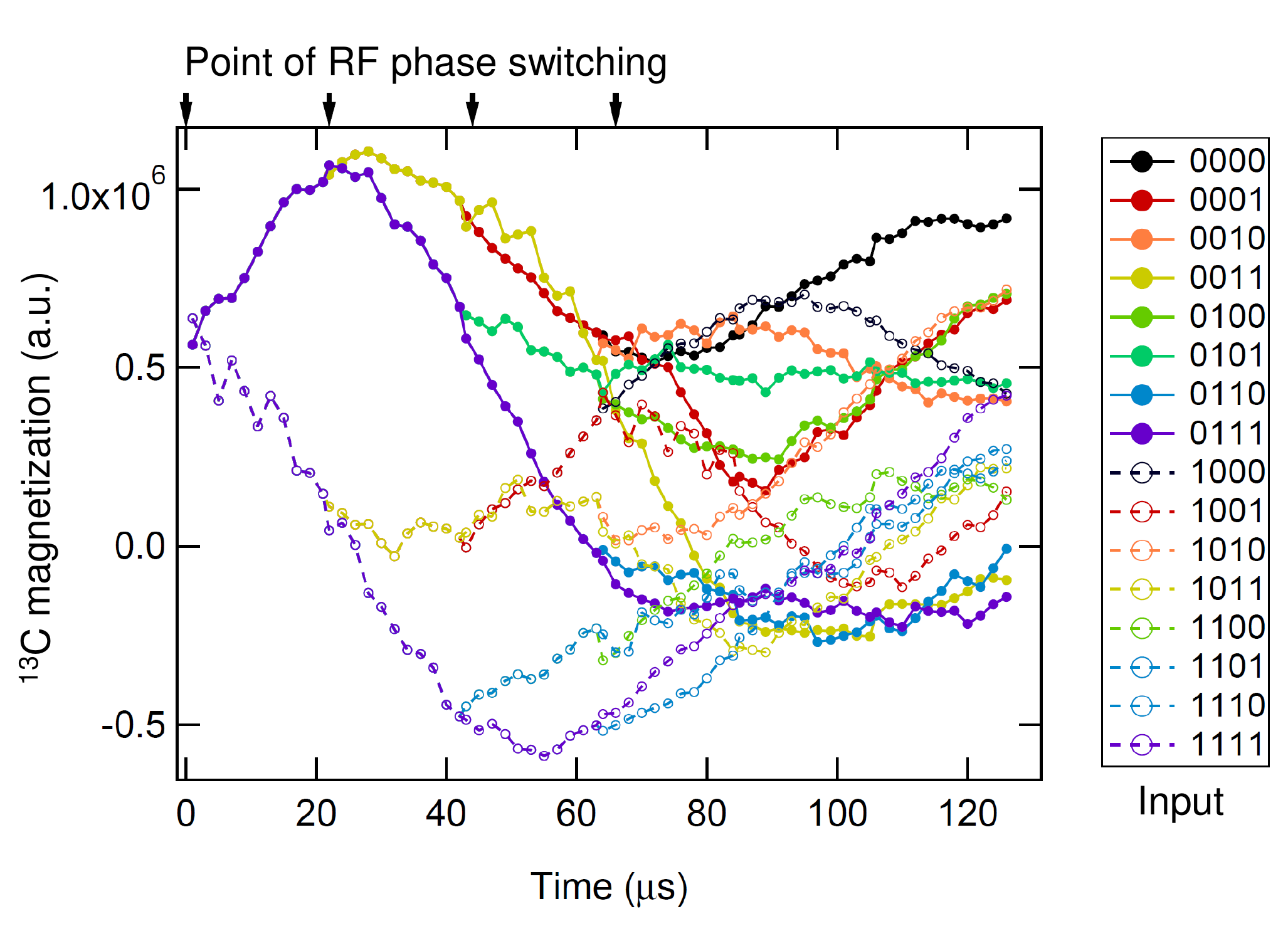}}
\caption{\label{fig1}
\small
Dynamics of the reservoir. The $Z$ component of the $^{13}$C spin in the interaction frame under RF irradiation, of which the phase is switched according to the input stream, is plotted.
}
\end{figure}

The quantum dynamics is now exploited for nonlinear machine learning tasks in this work.
While in Ref.~\cite{Fujii17}, quantum reservoir computing has been proposed to solve nonlinear temporal tasks, we here simplify the problem to non-temporal nonlinear tasks such as classification and pattern recognition to demonstrate computational capability of quantum dynamics for those tasks.
More precisely, we consider a learning problem of a nonlinear function $y=f({\bf s})$ using $K$ training data sets including input stream, $\{s_{l,k}\}_{l=1,k=1}^{L,K}$ ($s_{l,k} \in [0,1]$), where $L$ is the length of the stream, and corresponding teacher $\bar{y_k}$.
In order to feed an input stream to the reservoir, we globally rotate a part of the reservoir, the $^1$H spins, according to the input.
In the experiment shown in Fig.~2, we switched the RF phase to $\theta_{l,k}=\arccos(2 s_{l,k} -1)$.
The input is processed by $M$ cycles of the time evolution $U$, each of which is followed by the ensemble measurement of a part of the reservoir spin, the $^{13}$C spin.
We repeat $L$ series of the input process and the cycles of the evolution for each of $K$ instances of the input $\{ s_{l,k}\}$.
Thereby, we get the $LM$ signals $x_{l,m}^{(k)}$ in total for a given input instance $\{ s_{l,k}\}_{l=1}^{L}$ as shown in Fig.~2, where $M=11$ and $L=4$ and the sampling interval was 2 $\mu$s.

The output from the machine is defined as a weighted sum of the $LM$ signals $x_{l,m}^{(k)}$:
\begin{eqnarray}
y_k = \sum _{m=1}^{M} \sum _{l=1}^{L} W_{l,m} x_{l,m}^{(k)},
\end{eqnarray}
where $W$ is a $LM$-dimensional weight vector.
After taking signals with $K$ sets of the training data, which are denoted by $K \times LM$ matrix, $R$, $W$ is determined as to minimize the mean squared error, $\sum{(\bar{y_k}-y_k)^2}$.
The minimization is performed by taking the Moore-Penrose inverse of the matrix $R$.

First we analyze the computational capabilities in terms of two benchmark tasks under a binary sequential input (See also Supplementary Information).
One is the input recognition task, which measures how well each input is reconstructed from the quantum reservoir dynamics.
It shows that the input streams are embedded into the quantum reservoir. 
Another is the parity check task, which tests how well a nonlinear transformation can be done on the input sequence embedded into the reservoir.

The results are shown in Fig.~3 for different number of the sampling points $M=2,3,4,6,11$.
The more we increase the number of the sampling points, the higher performance we can obtain in the both cases, as numerically shown in Ref.~\cite{Fujii17}.
With $M=11$, almost all inputs, especially first three inputs, are nicely reconstructed from the quantum reservoir dynamics, which shows that the input streams are embedded into the quantum reservoir.
When the learned output, which is a real number, is rounded into either 0 or 1 with a threshold 0.5, then all the first three inputs are completely reconstructed and the final bit is reconstructed except for only two patterns out of 16.

The two bit parity of the 1st and 3rd inputs and that of the 2nd and 3rd inputs in the reservoir, that is, non-linear transformation of the input sequence embedded into the quantum reservoir can be obtained with low mean squared errors.
We further demonstrate generalization tasks of simple boolean functions, which are nonlinear functions when the binary information is embedded into a continuous variable.
The mean squared errors of multi-bit XOR (parity), NAND, 1 bit adder, and 2 bit adder functions are shown in Tab.~\ref{tab1}.
We also demonstrate the learning of various functions, $y = f(s_1, s_2)$, with not binary but continuous variable inputs: multiplication, $f(s_1, s_2)=s_1 \times s_2$, division, $f(s_1, s_2)=s_1/(1+s_2)$, and two nonlinear functions, $f(s_1, s_2)=s_1 s_2(1-s_1)$ (non-linear I) and $f(s_1, s_2)=s_1^2 + s_2^2$ (non-linear II).
We fed two inputs, $s_1$ and $s_2$, each in 0.125 increments from 0 to 1 into the system and obtain $8 \times 8 = 64$ input patterns in total.
By using them as training data, we obtained the output for each input, as shown in Fig.~4.
We can learn the functions by using {\it spatial multiplexing} technique~\cite{Nakajima18} with the low mean squared errors as shown in Tab.~\ref{tab1} (See also Supplementary Information).
It would be beneficial to investigate the generalization capability of the platform with different sampling intervals, molecules, and spatial multiplexing in the future work.

\begin{figure}
\center{\includegraphics[width=1.\linewidth]{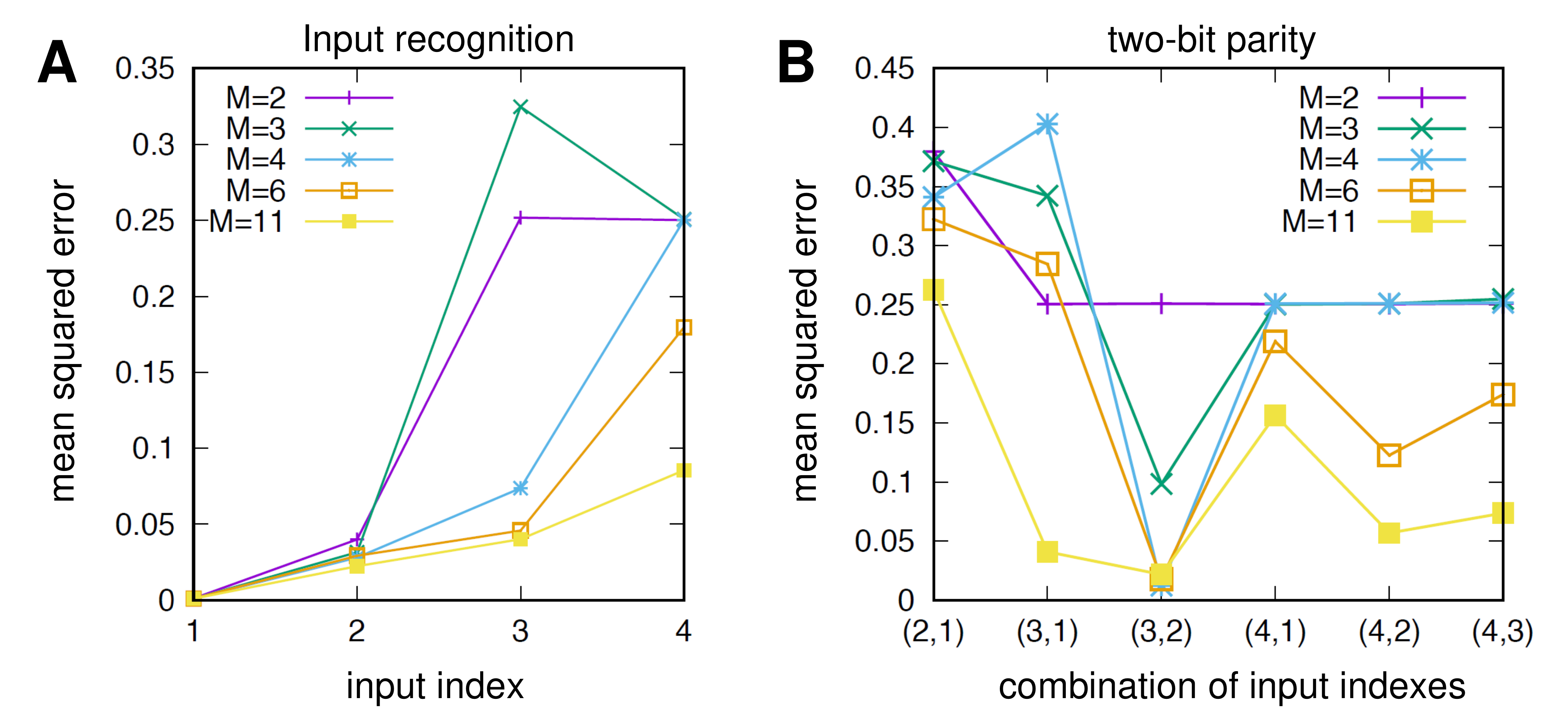}}%
\caption{Performance of input recognition (A) and two-bit parity check tasks (B) measured by the mean squared error.
Each plots correspond to the number of the sampling points $M=2,3,4,6,11$.
\label{fig4}}
\end{figure}

\begin{figure}
\center{\includegraphics[width=1.\linewidth]{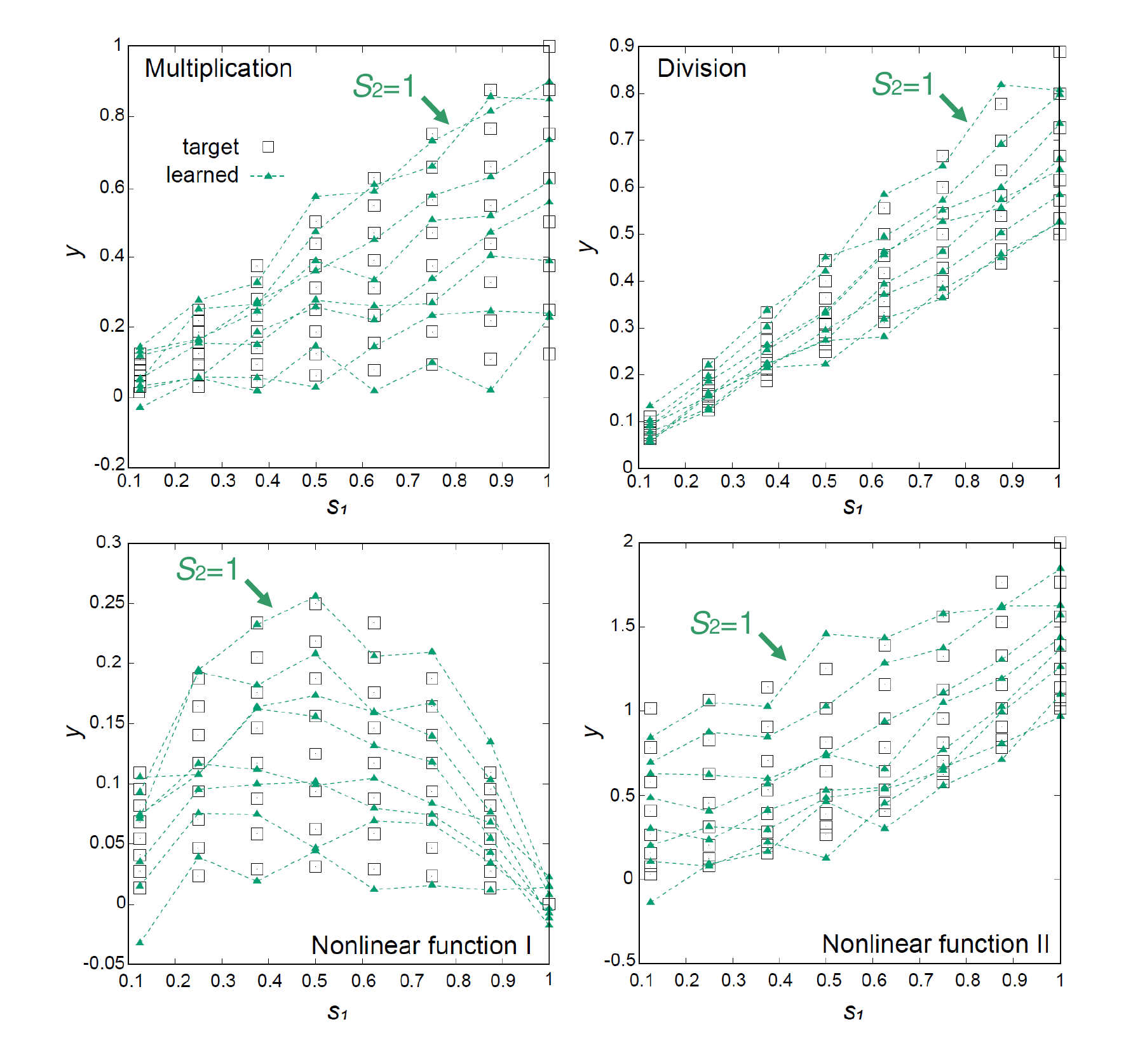}}%
\caption{Learned outputs of four functions with two continuous variable inputs.
\label{fig5}}
\end{figure}

\begin{table}[t]
\begin{center}
\begin{tabular}{c|c|c}
 \hline \hline
 & MSE & \# of error
\\
\hline \hline
2bit XOR  & $1.97\times 10^{-2}$ & 0
\\
\hline
3bit XOR  & $2.83\times 10^{-2}$ & 0 
\\
\hline
4bit XOR  & $1.99 \times 10^{-1}$ & 3
\\
\hline
NAND  & $2.26 \times 10^{-2}$  & 0
\\
\hline
1bit Adder 0th order & $2.26 \times 10^{-2} $& 0
\\
\hline
1bit Adder 1st order & $2.01 \times 10^{-2} $& 0
\\
\hline
2bit Adder 0st order & $4.36 \times 10^{-2}$& 1
\\
\hline
2bit Adder 1st order & $2.02 \times 10^{-1}$ & 4
\\
\hline
2bit Adder 2nd order & $1.44 \times 10^{-2}$& 0
\\
\hline
multiplication & $2.32 \times 10^{-3}$& -
\\
\hline
division & $5.22 \times 10^{-4}$& -
\\
\hline
non-linear I & $3.09 \times 10^{-4}$& -
\\
\hline
non-linear II & $7.64 \times 10^{-3}$& -
\\
\hline \hline
\end{tabular}
\caption{Performance of learning functions measured by the mean squared error and the number of erroneous outputs, where the output is digitized using a threshold 0.5.}
\label{tab1}
\end{center}
\end{table}

We discuss the total experimental duration for learning process.
It took 4 hours to get $4 \times 11 = 44$ values of the reservoir dynamics for every input stream.
Because the signal from the thermal polarization $\epsilon \sim 0.003\%$ was very small, we accumulated 16 times for acquiring each value with the interval of 20 sec for waiting the system to reach thermal equilibrium.
If the signal-to-noise ratio (SNR) were improved, we could reduce both the number of the accumulations and the signal acquisition window, which we set presently to 2 ms.
When we decrease the signal acquisition window by a factor of $R$, the SNR is decreased by a factor of $\sqrt{R}$.
If the signal acquisition window is sufficiently shorter than the coherence time, the reservoir state is less affected by the acquisition and therefore we can monitor several values at once, which is also important for temporal machine learning tasks, as mentioned later.
Thus, we could decrease the time of learning process by the factor of the square of the SNR improvement factor.
If the polarization were increased to 10\%, the SNR could be improved by the factor of 3,000.
We have already accomplished the nuclear polarization of $>$ 10\% in an isotopically-labelled crystal with repeated application of dynamic nuclear polarization (DNP) using photoexcited triplet electrons~\cite{Henstra90,Tateishi14} with the duration of 20 sec at room temperature~\cite{Negoro10}, where high fidelity control is compatible.
With such a high polarization, the 44 values for each input stream can be monitored with a sufficiently high SNR with one sequence within 20 sec.

In this work, we demonstrated non-temporal machine learning tasks with the globally controlled ensemble dynamics of partially-polarized spins well isolated in a solid.
If we want to apply for a temporal machine learning task~\cite{Fujii17}, long input stream should be fed into the reservoir and monitored with one sequence.
In addition, some form of initialization is necessary before every input.
For example, qubits can be initialized by DNP at very low temperatures $\sim 1$ K~\cite{deBoer74}.
The high fidelity control of a solid state spin ensemble has been already demonstrated albeit at room temperature~\cite{Ryan08}.
Development of the initialization technology compatible with high fidelity control~\cite{Cho07} is important for feeding a long input stream and realizing a temporal machine learning task, which makes the ensemble NMR quantum learning machine more powerful.
The other important approach to boost the computational power is increasing the number of the reservoir spins, as numerically demonstrated in Ref.~\cite{Fujii17}.
The technology for scaling up compatible with high fidelity control is within reach.
There are various oligomer and polymer with more than 50 globally controllable spins~\cite{Lloyd93}.
The simulation of the ensemble dynamics of more than 50 qubits with the initialized state as well as the partially-polarized state is computationally hard for classical computers.

Our implementation of the quantum reservoir and demonstration of nonlinear information processing in it paves a road toward exploiting quantum computational supremacy in NMR ensemble systems for information processing with reachable technologies.

\begin{acknowledgments}
M.N., K.F., and K.N. are supported by JST PRESTO Grant Number JPMJPR15E7, JPMJPR1668, and JPMJPR1666, Japan.
K.F. is supported by KAKENHI No.16H02211, JST ERATO Grant Number JPMJER1601, and JST CREST Grant Number JPMJCR1673.
K.N. is supported by KAKENHI No.16KT0019, No.15K16076, and No.26880010.
\end{acknowledgments}


\section{Supplementary Information}
\noindent{\it Methods ---}
The single crystal was grown from a deuterated water with 1-$^{13}$C-{\it l}-alanine and $^2$H$_7$-{\it l}-alanine with the weight ratio, 1:10.
Amino protons were replaced with deuterons in deuterated water.
The crystal size was 2 mm $\times$ 3 mm $\times$ 5 mm.
We used homebuilt spectrometer and NMR probe.
We accumulated FID (free induction decay) signals with the interval of 20 sec for waiting $^1$H spins to reach thermal equilibrium.
By several $\pi/2$ pulses to the $^{13}$C spin before the sequence shown in Fig.~1B, we saturated the $^{13}$C magnetization.

\noindent{\it Details of performance analysis ---}
To analyze the binary task performance, we experimentally acquired the quantum reservoir dynamics for two times for each input stream.
One is used for the learning process, and the other is for the evaluation of the performance.
To make the learning process stable to avoid overfitting, each experimental data for learning are replicated to $10^{4}$ data by adding a Gaussian noise with mean zero and variance $10^{4}$ (note that the maximal signal strength was $10^6$, as shown in Fig.~2).

To boost the learning performance for the functions with continuous variable inputs, we make use of spatial multiplexing technique~\cite{Nakajima18}.
We fed two inputs, $s^\prime_1$ and $s^\prime_2$, into the system each in 0.125 increments from $-1$ to 1 with a rotation gate with the angle, $\arccos s^\prime $, ranging from 0 to $\pi$.
We obtained $16 \times 16 = 256$ patterns of the dynamics.
For continuous variable tasks, we define the input of the function as $s_{l} = |s^\prime_{l}|$.
For example, for a functional input ($s_1$, $s_2$) = (0.25, 0.75), we utilize four patterns with the system input ($s^\prime_1$, $s^\prime_2$) = (0.25, 0.75), (-0.25, 0.75), (0.25, -0.75), (-0.25, -0.75).

In order to analyze the computational capability of the system, we examine two cases of learning schemes.
In the first case, shown in Fig.~4 and Tab.~I, the signals corresponding to an input are used as the evaluation data, and the signals collected for the remaining 63 inputs are used as the training data by augmenting the data further with addition of noise.
In the second case, all the sets of collected data are further augmented by adding noise as the training data and the original data without noise are used as the evaluation data set.
The mean squared errors are 4.53 $\times$ 10$^{-5}$, 8.40 $\times$ 10$^{-6}$, 5.84 $\times$ 10$^{-6}$, and 1.42 $\times$ 10$^{-4}$ for multiplication, division, nonlinear function I, and II, respectively, which are much smaller than those in the first case.

\end{document}